\documentstyle[amssymb,aps]{revtex}

\begin{document}
\title{{\bf Resonances and spectral properties of detuned OPO pumped by fluctuating
sources}}
\author{A. Porzio$^{\dagger }$\thanks{%
Tel:+39 081 676188, Fax:+39 081 676346, email: alberto.porzio@na.infn.it},
C. Altucci$^{{\times }\dagger }$, P. Aniello$^{\ddagger }$, C. de Lisio$%
^{\dagger }$, and S. Solimeno$^{\dagger }$}
\address{Dip. Scienze Fisiche, Univ. ''Federico II''\\
$^{\dagger }$INFM Unit\`{a} di Napoli,\\
$^{\ddagger }$INFN Sezione di Napoli,\\
Compl. Univ. Monte Sant'Angelo, via Cintia, I-80126 Napoli, Italy\\
${\times }$Dip. di Chimica,Univ. della Basilicata, Potenza, Italy}
\maketitle

\begin{abstract}
Twin beam fluctuations are analyzed for detuned and mismatched OPO
configurations. Resonances and frequency responses to the quantum noise
sources (quantum and pump amplitude/phase fluctuations) are examined as
functions of cavity decay rates, excitation parameter and detuning. The
dependence of self- and mutual correlations of beam amplitudes and phases on
detuning, mismatch and damping parameters is discussed.
\end{abstract}

\bigskip

generation, frequency conversion

\section{\bf Introduction}

Continuous wave twin beams owe their popularity to the mutual quantum
correlations. It has been shown theoretically (see f.i. \cite
{Reynaud,Drummond,Milburn,Collett,Yurke,CollettI,Savage,CollettII}) and
proved experimentally (see f.i. \cite{Heidmann,ReynaudI}) that this feature
can be exploited for replacing classical fields in shot noise limited
optical measurements. In particular, the spectrum difference of the two
beams is shaped like a Lorentzian with a rather good noise suppression at
zero frequency. On the other hand, Fabre et al. \cite{FabreI} have called
the attention on the squeezing exhibited by each beam for pump intensity
larger than four times the threshold value. They have also shown that
additional squeezing can be imposed on each beam by using the photocurrent
detected with the other one.

These properties are contrasted by unsuited damping coefficients of the OPO
cavity, imperfect tuning of the three mode resonances, limited intensity and
amplitude/phase fluctuations of the laser pump. First Lane et al. \cite{Lane}
drew the attention on the critical dependence of the noise suppression in
the difference spectrum on the cavity mismatch, as experimentally confirmed
by several authors (see f.i. \cite{Nabors,Wong}). Stable and narrow line
lasers emit in general low intensity beams. Then, for observing single beam
squeezing it is essential to lower the threshold by reducing the pump mode
decay rate. This solution makes the OPO a triply resonant device featuring
relaxation oscillations \cite{Bjork,Lee,Porzio}. Due to the unavoidable OPO
crystal losses the amplitude/phase fluctuations of the pump contribute to
the difference spectral noise at low frequency, where quantum noise
suppression is expected. The situation becomes worse and worse as the cavity
is detuned from the resonant configuration.

These few remarks point to the several parameters which the features of an
OPO depend on, namely, cavity damping coefficients, mismatch of these
parameters, degree of excitation above threshold, crystal absorption,
deviation from resonance condition (detuning), and pump amplitude/phase
fluctuations. The last ones play a prominent role in the lower part of the
frequency spectra of all relevant quantities. Many papers have been
published discussing the role of some of these parameters. Here we aim to
discuss analytically the twin beam fluctuations by encompassing
systematically the many parameters mentioned above.

The representation of the three modes (signal/idler $k=1$, $2$, pump $k=0$)
in the form $a_{k}=r_{k}e^{-i\phi _{k}}\left( 1+\mu _{k}\right) $ lends
itself to separate the steady state amplitude $r_{k}$ from the fluctuating
relative amplitude $\mu _{k}$ and phase $\phi _{k}.$ In regard to the
fluctuations, the OPO behaves as a forced linear system. Then, the Fourier
transforms $\hat{a}_{k}\left( \omega \right) $ of the combination $\delta
a_{k}=\mu _{k}-i\phi _{k}$ and the adjoint $\hat{a}_{k}^{\ddagger }\left(
\omega \right) \equiv \hat{a}_{k}^{\dagger }\left( -\omega ^{\ast }\right) $
form an algebraic system. The dependence of the $\hat{a}_{k}\left( \omega
\right) $ on the noise sources is embodied in the frequency responses $%
K_{kl}^{0,\frac{\pi }{2}}\left( \omega \right) $ of the $k$-th mode to the $%
l $-th noise source. These responses are analytic functions of $\omega $
with, all together, six complex poles (system resonances) in the half-plane $%
\mathop{\rm Im}%
\left( \omega \right) \geq 0$. Their location (generally complex or purely
imaginary)\ influences the field correlations in the time domain and, hence,
the relative frequency spectra. For example, when the poles have large real
parts the system exhibits relaxation oscillations, responsible of typical
peaks in photocurrent spectra. In short the $K_{kl}^{0,\frac{\pi }{2}}\left(
\omega \right) $ and their poles provide a link among the OPO, pump laser
parameters and the statistical features of the twin beams.

While several authors have shown that the unavoidable cavity mismatch
reduces the mode correlation, less attention has been paid to the effects of
detuning, apart from the analysis of Ref. \cite{FabreI} dedicated to OPO
with pump bandwidths much larger than the signal/idler ones, i.e. in the
adiabatic limit. For large detunings bistability, self--pulsing and chaos
may appear \cite{GrahamIV,Lugiato}, while for moderate deviations from
cavity resonance, modifications of the spectra of the single beams and their
differences are expected. In presence of detuning the decay constants become
complex, $\kappa _{k}=\gamma _{k}\;+i\varphi _{k}$, and the equations of
motion of $\mu _{k}$ do not separate from those of $\phi _{k}$. In
particular, the pump phase diffusion leaks into the single mode amplitude
spectrum by enhancing the disturbing effects of the classical pump
fluctuations. From the analytic point of view, the detuning has the effect
of breaking the symmetry of the frequency responses, $K_{kl}^{0,\frac{\pi }{2%
}}\left( \omega \right) \neq K_{kl}^{0,\frac{\pi }{2}*}\left( -\omega
\right) $, so that $\mu _{k}\left( t\right) $ and $\phi _{k}\left( t\right) $
do not commute with the respective quantities at different times.

The model, herein discussed, is an extension of the quantum analysis of
Graham and Haken \cite{Graham,GrahamI} and Fabre et al. \cite{FabreI}. In
previous works the variance and covariance of the OPO beams have been
analyzed by representing the fields either as operators \cite{Bjork} or
classical variables. While Reid and Drummond \cite{Reid} have taken in full
account the fluctuating phase factors $e^{-i\phi _{k}}$ by using the
stochastic equation method, the operator representation has been limited to
models neglecting the phase diffusion processes. To fill the gap between
these two approaches, quantum Langevin equations will be used for
calculating second and fourth order correlations. The operator
representation offers the advantage of introducing the commutators of the
field variables at different times. Some of these commutators vanish at
resonance, so that their Fourier transforms provide an additional signature
of the detuning. Essential to these calculations will be the assumption of
Gaussian statistics of the mode operators. Most of the numerical examples
discussed below refer to OPO exhibiting relaxation oscillations, that is
having pump decay rates comparable with the signal/idler ones.

The paper is organized as follows. In Sec. 2 the basic equations for an OPO
are reviewed. The steady state solutions are recalled in Sec. 3, while in
Sec. 4 it is introduced the linearized set of the equations of motion for
the fluctuating parts of the internal modes. Then, switching in Sec. 5 to
the frequency domain, the Fourier transforms of the fluctuations are
expressed as functions of the noise inputs. Moreover, the quadratures of the
single beams are represented by combinations of the quantum and classical
(pump amplitude and phase) noise sources times suitable transfer functions.
The system resonances are discussed in Sec. 6 by dwelling on the dependence
of the pole location on the cavity decay constants and excitation parameter.
Eventually, the photocurrent fluctuation spectra are examined in Sec. 8 with
the help of the transfer functions $K_{kl}^{0,\frac{\pi }{2}}$ discussed in
Sec. 7.

Details on the solution of the algebraic system describing the fluctuations
of the OPO modes in the Fourier domain are given in Appendix A and B. The
damping coefficients of the OPO cavity and the losses of the nonlinear
crystal have been indicated respectively by $\gamma _{k}$ and $\varkappa
_{k} $, by adding an apex for the total losses $\gamma _{k}^{\prime }=\gamma
_{k}+\varkappa _{k}$. Quantities normalized with respect to the average
damping $\gamma =\left( \gamma _{1}^{\prime }+\gamma _{2}^{\prime }\right)
/2 $ have been represented by superposing a tilde. $\delta _{1}=\left(
\gamma _{1}^{\prime }-\gamma _{2}^{\prime }\right) /2=-\delta _{2}$ measures
the cavity mismatch.

\section{\bf Equations of motion}

The evolution of the mode amplitudes $a_{0}$ (pump), $a_{1}$ and $a_{2}$
(signal/idler) of an OPO is described by a system of Langevin equations 
\begin{equation}
\frac{\partial }{\partial t}\left( 
\begin{array}{c}
a_{1} \\ 
a_{2} \\ 
a_{0}
\end{array}
\right) =\left( 
\begin{array}{c}
-\kappa _{1}^{\prime }a_{1}+2\chi a_{0}a_{2}^{\dagger }+R_{1}^{\prime } \\ 
-\kappa _{2}^{\prime }a_{2}+2\chi a_{0}a_{1}^{\dagger }+R_{2}^{\prime } \\ 
-\kappa _{0}^{\prime }a_{0}-2\chi ^{*}a_{1}a_{2}+\epsilon +R_{0}^{\prime }
\end{array}
\right)  \label{Langevin}
\end{equation}
with $R_{k}^{\prime }\left( t\right) $ independent fluctuating
delta-correlated Langevin forces and $\epsilon =\left| \epsilon \right|
e^{-i\phi _{p}}$ with the pump phase $\phi _{p}\left( t\right) $ undergoing
a slow diffusion, $\left\langle \left( \phi _{p}\left( \tau \right) -\phi
_{p}\left( 0\right) \right) ^{2}\right\rangle =2\Delta \nu _{L}\tau .$ The
coefficients $\kappa _{k}^{\prime }$ ($k=0,1,2$) stand for 
\begin{equation}
\kappa _{k}^{\prime }=\gamma _{k}+\varkappa _{k}+i\varphi _{k}=\gamma
_{k}^{\prime }+i\varphi _{k}=\kappa _{k}+\varkappa _{k}  \label{damping}
\end{equation}
with $\gamma _{k}$ and $\varkappa _{k}$ the cavity mode damping rates,
associated respectively with the output mirror and the other loss mechanisms
(mainly due to the non--linear crystal), $\varphi _{k}$ the detunings and $%
2\chi $ the strength of the parametric interaction.

The operators $R_{k}^{\prime }$ are proportional to the vacuum fluctuation $%
b_{in_{k},1}$ entering the system through the output mirror and to the noise 
$b_{in_{k},2}$ associated with the crystal losses, 
\begin{equation}
R_{k}^{\prime }=\sqrt{2\gamma _{k}}b_{in_{k},1}+\sqrt{2\varkappa _{k}}%
b_{in_{k},2}\equiv R_{k}+\sqrt{2\varkappa _{k}}b_{in_{k},2}
\label{input-output}
\end{equation}

Applying now the time reversal to the above Langevin equations \cite
{Collett,Gardiner} it can be easily shown that output, internal and input
fields $\left( b_{out_{k},1},a_{k},b_{in_{k},1}\right) $ are connected in
presence of detuning by the standard boundary conditions \cite{FabreI} 
\begin{equation}
b_{out_{k},1}=\sqrt{2\gamma _{k}}a_{k}-b_{in_{k},1}  \label{boundary}
\end{equation}

Finally, the amplitude-phase representation 
\begin{equation}
a_{k}=r_{k}e^{-i\phi _{k}}\left( 1+\mu _{k}\right)  \label{amplitude}
\end{equation}
will be adopted in the following, with $r_{k}$ the steady state amplitude, $%
\phi _{k}$ the phase, and $\mu _{k}$ the relative amplitude fluctuation.

\section{S{\bf teady-state solution}}

Dropping the Langevin forces and the fluctuating amplitudes, the system (\ref
{Langevin}) admits the steady state solution (see Eq. (\ref{amplitude})) 
\[
\alpha _{k}=\left\langle a_{k}\right\rangle =r_{k}e^{-i\phi _{k}} 
\]
($\alpha _{k}^{*}=\left\langle a_{k}^{\dagger }\right\rangle $). Although $%
\phi _{k}$ fluctuates it has been included in order to get average values of
some phase combinations.

The signal/idler amplitudes $\alpha _{j}$ $(j=1,2)$ are different from zero
for $\left| \epsilon \right| $ above a threshold 
\begin{equation}
\epsilon ^{th}=\frac{\left| \kappa _{0}^{\prime }\right| \sqrt{\left| \kappa
_{1}^{\prime }\kappa _{2}^{\prime }\right| }}{2\chi }=\frac{\gamma
_{0}^{\prime }\sqrt{\gamma _{1}^{\prime }\gamma _{2}^{\prime }}}{2\chi \cos
\psi }=\left| \kappa _{0}^{\prime }\right| r_{0}  \label{threshold}
\end{equation}
only if the relative detunings $\varphi _{j}$ (see Eq. (\ref{damping})) are
proportional to the damping factors $\gamma _{j}^{\prime }=\gamma _{j}+$ $%
\varkappa _{j}$ of the respective modes, i.e. 
\[
\frac{\varphi _{1}}{\gamma _{1}^{\prime }}=\frac{\varphi _{2}}{\gamma
_{2}^{\prime }}=\tan \psi \;,\quad \frac{\varphi _{0}}{\gamma _{0}^{\prime }}%
=\tan \left( \psi _{0}-\psi \right) 
\]
Most OPO cavities are stabilized on the pump mode, so that in the following
we will put $\psi =\psi _{0}.$

Amplitudes $r_{k}$ and phases $\phi _{k}$ of the three modes result
interconnected 
\begin{eqnarray}
\left| \kappa _{1}^{\prime }\right| r_{1}^{2} &=&\left| \kappa _{2}^{\prime
}\right| r_{2}^{2}=\left( {\cal E}-1\right) \left| \kappa _{0}^{\prime
}\right| r_{0}^{2}  \nonumber \\
\left\langle \phi _{1}+\phi _{2}-\phi _{0}\right\rangle &=&\psi
\label{phases-I}
\end{eqnarray}
with ${\cal E}$ related to the amplitude parameter $E=\left| \epsilon
/\epsilon ^{th}\right| $ by 
\begin{equation}
{\cal E}=\sqrt{E^{2}-\sin ^{2}\psi }+1-\cos \psi  \label{Epsilon}
\end{equation}
In absence of detuning ${\cal E}$ reduces to $E$. Notice that the relation
between the frozen phases of the three beams (see f.i. Eq. (2.9) of Ref. 
\cite{Lane}) depends on the detuning through the phase $\psi $.

In turn $r_{0}$ and $\,\phi _{0}$ are respectively given by 
\begin{eqnarray}
r_{0} &=&\frac{\sqrt{\left| \kappa _{1}^{\prime }\kappa _{2}^{\prime
}\right| }}{2\chi }  \nonumber \\
\left\langle \,\phi _{p}-\phi _{0}\right\rangle &=&\psi -\psi _{p}
\label{phases-II}
\end{eqnarray}
with $\phi _{p}$ the phase of the external pump and $\psi _{p}=\arcsin
\left( \sin \psi /E\right) $.

In the following the parameter 
\[
C^{2}=\left( {\cal E}-1\right) \left| \kappa _{0}^{\prime }\right| r_{0}^{2} 
\]
will be used.

\section{\bf Fluctuations}

The phase relations obtained for the steady state solutions are satisfied
only approximately. In particular, the phases $\phi _{1}+\phi _{2}-\phi _{0}$
and $\phi _{0}-\phi _{p}$ (Eqs. (\ref{phases-I}--b) and (\ref{phases-II}-b))
undergo stable fluctuations, while $\phi _{1}-\phi _{2}$ follows an undamped
diffusion process. On the other hand, \cite{Graham} 
\begin{equation}
\dot{a}_{k}=r_{k}e^{-i\phi _{k}}\delta \dot{a}_{k}  \label{derivative}
\end{equation}
where $\delta a_{k}$ $=\mu _{k}-i\phi _{k}$ (Eq. (\ref{amplitude}))

By exploiting the smallness of the quantities $\phi _{1}+\phi _{2}-\phi
_{0}-\psi $ and $\phi _{0}-\phi _{p}+\psi -\psi _{p}$ the system (\ref
{Langevin}) can be linearized with respect to the fluctuating phases $\phi
_{k}$ and amplitudes $\mu _{k}$ \cite{Graham}. Next, changing the phases $%
\phi _{k}$ and $\phi _{p}$ into $\phi _{j}^{\prime }=\phi _{j}-\frac{1}{2}%
\psi _{p}$, $\phi _{0}^{\prime }=\phi _{0}+\psi -\psi _{p}$, and$\;\phi
_{p}^{\prime }=\phi _{p}+\psi _{p}$ does not change the statistical
properties of the system solutions. Thus, keeping the old symbols for the
new quantities,.the linearized system reads: 
\begin{eqnarray}
\frac{\delta \dot{a}_{1}}{\kappa _{1}^{\prime }}+\delta a_{1}-\delta
a_{2}-\delta a_{0} &=&Z_{1}^{\prime }  \nonumber \\
\frac{\delta \dot{a}_{2}}{\kappa _{2}^{\prime }}+\delta a_{2}-\delta
a_{0}-\delta a_{1} &=&Z_{2}^{\prime }  \nonumber \\
\frac{e^{i\psi _{0}}}{{\cal E}-1}\left( \frac{\delta \dot{a}_{0}}{\kappa
_{0}^{\prime }}+\delta a_{0}\right) +\delta a_{1}+\delta a_{2}
&=&Z_{0}^{\prime }+Z_{\epsilon }+Z_{\phi }  \label{TRO}
\end{eqnarray}
with 
\begin{eqnarray}
Z_{j}^{\prime } &=&\frac{e^{i\phi _{j}}R_{j}^{\prime }}{\kappa _{j}^{\prime
}r_{j}}\qquad (j=1,2)  \nonumber \\
Z_{0}^{\prime } &=&\frac{1}{{\cal E}-1}e^{i\psi _{0}}\frac{e^{i\phi
_{0}}R_{0}^{\prime }}{\kappa _{0}^{\prime }r_{0}}  \nonumber \\
Z_{\epsilon } &=&\frac{E}{{\cal E}-1}e^{i\psi _{p}}\mu _{\epsilon } 
\nonumber \\
Z_{\phi } &=&-i\frac{E}{{\cal E}-1}e^{i\psi _{p}}\phi _{p}  \label{Z}
\end{eqnarray}
having denoted by $\mu _{\epsilon }=\delta \epsilon /\epsilon $ and $\phi
_{p}$ the pump relative excess noise and phase fluctuation respectively.

The output field, having average intensity $\bar{I}_{j}=2\gamma
_{j}r_{j}^{2}=\left( 2\gamma _{j}/\gamma _{j}^{\prime }\right) \cos \psi
\;C^{2}$, is given by (see Eq.(\ref{boundary})) 
\begin{equation}
b_{out_{j}}=\sqrt{2\gamma _{j}}r_{j}e^{-i\phi _{j}}\left( 1+\mu _{j}-\frac{%
\kappa _{j}^{\prime }}{2\gamma _{j}}Z_{j}\right)  \label{field-out}
\end{equation}
where $Z_{j}$ is given by (\ref{Z}-a) with $R_{j}^{\prime }$ replaced by $%
R_{j}.$

Before concluding this section, it is worth looking at the approximations
underlying the linear system (\ref{TRO}). Products of phase-factors $%
e^{i\phi }e^{-i\chi }$ can be represented by the generalized
Campbell-Hausdorff formula (see f.i. \cite{Saltinger} p. 35) 
\[
e^{i\phi }e^{-i\chi }=\exp \left( i\phi -i\chi +\frac{1}{2}\left[ \phi ,\chi %
\right] +\frac{i}{12}\left[ \left[ \phi ,\chi \right] ,\phi +\chi \right]
\cdots \right) 
\]
so that choosing $\phi =\phi \left( t\right) $,$\ \chi =\phi \left(
t+dt\right) $ we have for the derivative of $a_{k}\left( t\right) $%
\[
\dot{a}_{k}\left( t\right) =r_{k}e^{-i\phi _{k}}\left( \delta \dot{a}%
_{k}\left( t\right) -\frac{1}{2}\left[ \phi _{k}\left( t\right) ,\dot{\phi}%
_{k}\left( t\right) \right] +\frac{i}{6}\left[ \left[ \phi _{k},\dot{\phi}%
_{k}\right] ,\phi _{k}\right] \cdots \right) 
\]
which reduces to Eq. (\ref{derivative}) at the first order in the
fluctuating quantities.

\section{\bf Frequency analysis}

Now it is worth introducing the Fourier transforms 
\begin{equation}
{\delta a_{k}\left( t\right)  \choose \delta a_{k}^{\dagger }\left( t\right) }%
=\frac{1}{\sqrt{2\pi }}\int_{-\infty }^{\infty }e^{i\omega t}%
{\hat{a}_{k}\left( \omega \right)  \choose \hat{a}_{k}^{\ddagger }\left( \omega \right) }%
d\omega \;  \label{Fourier-transform}
\end{equation}
with $\hat{a}_{k}^{\ddagger }\left( \omega \right) =\hat{a}_{k}^{\dagger
}\left( -\omega ^{*}\right) $ the adjoint of the Fourier component relative
to the frequency $-\omega $. In the following we will consider generally
complex frequencies $\omega $ and the symbol $\ddagger $, operating on the
product of a c-function times an operator, will indicate the transformation $%
\left[ f\left( \omega \right) \hat{a}\left( \omega \right) \right]
^{\ddagger }=f^{*}\left( -\omega ^{*}\right) \hat{a}^{\dagger }\left(
-\omega ^{*}\right) $.

In the frequency domain Eq. (\ref{TRO}) reduces to the algebraic system: 
\begin{equation}
\begin{array}{ll}
\Delta _{1}\hat{a}_{1}-\hat{a}_{0}-\hat{a}_{2}^{\ddagger } & =\hat{Z}%
_{1}^{\prime } \\ 
\Delta _{2}\hat{a}_{2}-\hat{a}_{0}-\hat{a}_{1}^{\ddagger } & =\hat{Z}%
_{2}^{\prime } \\ 
\Delta _{0}\hat{a}_{0}+\hat{a}_{1}+\hat{a}_{2} & =\hat{Z}_{0}^{\prime }+\hat{%
Z}_{\epsilon }+\hat{Z}_{\phi }
\end{array}
\label{TRO-final}
\end{equation}
with 
\begin{equation}
\Delta _{j}=1+i\frac{\omega }{\kappa _{j}^{\prime }}\;,\qquad \Delta
_{0}=\left( 1+i\frac{\omega }{\kappa _{0}^{\prime }}\right) \frac{e^{i\psi
_{0}}}{{\cal E}-1}  \label{Delta-functions}
\end{equation}
According to Eq. (\ref{Z}-c,d) $\hat{Z}_{\epsilon }=\frac{E}{{\cal E}-1}%
e^{i\psi _{p}}\hat{\mu}_{\epsilon }$ and $\hat{Z}_{\phi }=-i\frac{E}{{\cal E}%
-1}e^{i\psi _{p}}\hat{\phi}_{p}$ depend respectively on the pump amplitude
excess noise $\hat{\mu}_{\epsilon }$ and on the phase $\hat{\phi}_{p}$.

Next, completing the system (\ref{TRO-final}) with the respective $\ddagger $%
-transformed equations we obtain for the Fourier transforms $\hat{a}%
_{k}\left( \omega \right) $, $\,$and $\hat{a}_{k}^{\ddagger }\left( \omega
\right) $ a sixth-order algebraic system depending on the forcing terms $%
\hat{Z}_{k}^{\prime }$, $\hat{Z}_{\epsilon }$, $\hat{Z}_{\phi }$, $\hat{Z}%
_{k}^{\prime \ddagger }$, $\hat{Z}_{\epsilon }^{\ddagger }$, $\hat{Z}_{\phi
}^{\ddagger }.$ Being an OPO a stable system (see f.i. Ref. \cite{Lane}), $%
\hat{a}_{k}\left( \omega \right) $, $\,\hat{a}_{k}^{\ddagger }\left( \omega
\right) $ may become singular only in the upper imaginary plane ($%
\mathop{\rm Im}%
\left( \omega \right) \geq 0$).

Now, solving the system (\ref{TRO-final}) completed by the $\ddagger $%
-counterpart yields 
\[
\hat{a}_{k}=\frac{A_{k}^{\ddagger }C_{k}-B_{k}C_{k}^{\ddagger }}{%
A_{k}A_{k}^{\ddagger }-B_{k}B_{k}^{\ddagger }} 
\]
with the functions $A_{k}\left( \omega \right) ,B_{k}\left( \omega \right)
,C_{k}\left( \omega \right) $ defined in Appendix A.

Next, introducing the Fourier transformed quadratures of the output modes $%
\hat{\mu}_{\beta _{k}}=\frac{1}{2}\left( \hat{\beta}_{k}+\hat{\beta}%
_{k}^{\ddagger }\right) $, and $\hat{\phi}_{\beta _{k}}=\frac{i}{2}\left( 
\hat{\beta}_{k}-\hat{\beta}_{k}^{\ddagger }\right) $ we have (see Appendix
A) 
\begin{eqnarray}
\hat{\mu}_{\beta _{j}} &=&\left( K_{jj}^{0}-\frac{\kappa _{j}^{\prime }}{%
2\gamma _{j}}\right) \otimes \hat{X}_{j}^{0}+K_{jj}^{0}\otimes \hat{X}%
_{\varkappa _{j}}^{0}+K_{jj^{\prime }}^{0}\otimes \hat{X}_{j^{\prime
}}^{\prime 0}+K_{j0}^{0}\otimes \left( \hat{X}_{0}^{\prime \psi _{0}}+\frac{E%
}{{\cal E}-1}\hat{X}_{\epsilon }^{\psi _{p}}-\frac{E}{{\cal E}-1}\hat{X}%
_{\phi _{p}}^{\frac{\pi }{2}+\psi _{p}}\right)  \nonumber \\
\hat{\phi}_{\beta _{j}} &=&i\left( K_{jj}^{\frac{\pi }{2}}-\frac{\kappa
_{j}^{\prime }}{2\gamma _{j}}\right) \otimes \hat{X}_{j}^{0}+iK_{jj}^{\frac{%
\pi }{2}}\otimes \hat{X}_{\varkappa _{j}}^{0}+iK_{jj^{\prime }}^{\frac{\pi }{%
2}}\otimes \hat{X}_{j^{\prime }}^{\prime 0}+iK_{j0}^{\frac{\pi }{2}}\otimes
\left( \hat{X}_{0}^{\prime \psi _{0}}+\frac{E}{{\cal E}-1}\hat{X}_{\epsilon
}^{\psi _{p}}-\frac{E}{{\cal E}-1}\hat{X}_{\phi _{p}}^{\frac{\pi }{2}+\psi
_{p}}\right)  \nonumber \\
&&  \label{quadratures}
\end{eqnarray}
where $K\otimes \hat{X}$ is a shorthand for $\frac{1}{2}\left( K\left(
\omega \right) b\left( \omega \right) +K^{\ast }\left( -\omega ^{\ast
}\right) b^{\ddagger }\left( \omega \right) \right) $.

The functions $K_{kl}^{0,\frac{\pi }{2}}$ represent the frequency responses
to the quantum noise sources and the classical phase and amplitude
fluctuations of the pump laser. They are analytic in \ $\omega $ with poles
in the upper half plane $%
\mathop{\rm Im}%
\left( \omega \right) \geq 0$ coincident with some zeros of $%
A_{k}A_{k}^{\ddagger }-B_{k}^{2}$, as shown in the following sections.

According to (\ref{quadratures}) the amplitudes and phases depend on the
quadratures $X_{k}^{\theta }=\frac{1}{2}\left( e^{i\theta }Z_{k}+e^{-i\theta
}Z_{k}^{\dagger }\right) $ of the $k$-th vacuum noise entering the cavity
through the coupling mirror and that relative to the crystal loss $\varkappa
_{k},$ $X_{\varkappa _{k}}^{\theta }$. The fluctuating amplitude $\mu
_{\epsilon }$ and phase $\phi _{p}$ of the external pump are represented
respectively by $X_{\epsilon }^{\theta }$ and $X_{\phi _{p}}^{\theta }$.

\section{\bf Resonances}

Important features of the OPO fluctuations can be investigated by extending
the system (\ref{TRO-final}) to complex values of $\omega $. This will be
done by retaining the above definition of the operator $\ddagger $
transforming an analytic function of $\omega $, with poles $\Omega _{r}$, in
a new one with poles $-\Omega _{r}^{\ast }$ and complex conjugate residues.
The system of six unknown functions $\hat{a}_{k}\left( \omega \right) $, $%
\hat{a}_{k}^{\ddagger }\left( \omega \right) $ is characterized by the sixth
order characteristic polynomial $D\left( \tilde{\omega}\right) =D^{\prime
}\left( \tilde{\omega}\right) \tilde{\omega}$ with a zero at the origin, 
\begin{equation}
D^{\prime }\left( \tilde{\omega}\right) =\tilde{\omega}^{5}+D^{\left(
5\right) }\tilde{\omega}^{4}+D^{\left( 4\right) }\tilde{\omega}%
^{3}+D^{\left( 3\right) }\tilde{\omega}^{2}+D^{\left( 2\right) }\tilde{\omega%
}+D^{\left( 1\right) }  \label{D-prime}
\end{equation}
with $\tilde{\omega}$ the frequency normalized to $\gamma .$ The
coefficients $D^{\left( j\right) }$ are reported in Appendix B (Eq.(\ref
{coefficients})).

For a detuned balanced cavity ($\delta =\left| \delta _{1,2}\right| =0$) $%
D^{\left( i\right) }$ are independent of $\tan \psi $ so that $D^{\prime }$
coincides with that of a resonant OPO having an effective excitation
parameter $E_{eff}=1+\left( {\cal E}-1\right) /\cos \psi $. In particular, $%
D^{\prime }$ factorizes into the product $D_{+}D_{-}$ of polynomials
discussed in the following for the resonant case. This means that a detuned
balanced OPO cannot be distinguished from a perfectly resonant one with an
appropriate pump strength.

The six zeros $\tilde{\Omega}_{r}$ $\left( r=0,\ldots ,5\right) $ of $%
D\left( \tilde{\omega}\right) $ represent the system resonances. In
particular, the real parts of their values indicate the oscillation
frequencies, whereas the imaginary ones stand for the corresponding damping
parameters, i.e. the resonance widths. Since $D\left( \tilde{\omega}\right)
=D^{\ddagger }\left( \tilde{\omega}\right) $, each resonance $\tilde{\Omega}%
_{r}$ having a finite real part is mirrored by $-\tilde{\Omega}_{r}^{\ast }$%
. Apart from the one at the origin $\left( \tilde{\Omega}_{0}=0\right) $,
the resonance$\ \tilde{\Omega}_{1}=iw_{1}$ is located on the positive
imaginary axis, while the remaining four ones may be either purely imaginary
or may have finite real parts. In fact, for excitation parameter varying in
an interval depending on the cavity damping factors, one or both couples $%
\tilde{\Omega}_{r}$, $-\tilde{\Omega}_{r}^{\ast }$ are replaced by two
purely imaginary roots. In the case of non--vanishing real part$\ \tilde{%
\Omega}_{2}=-\tilde{\Omega}_{3}^{\ast }=\tilde{\omega}_{2}+iw_{2}$ and $%
\tilde{\Omega}_{4}=-\tilde{\Omega}_{5}^{\ast }=\tilde{\omega}_{4}+iw_{4}$;
the system is characterized by two dampings $w_{2},w_{4}$, and frequencies $%
\tilde{\omega}_{2},\tilde{\omega}_{4}$. In particular, $w_{1}+2w_{2}+2w_{4}$
is independent of the detuning, whereas for a balanced OPO $w_{1}=2$.

The expressions of $\tilde{\Omega}_{r}$ are rather lengthy and complicated,
thus, for the sake of simplicity, we briefly discuss their major features,
and dwell on two noticeable cases. Fig. 1 displays the dependence of the
real and imaginary parts of $\tilde{\Omega}_{r}$ on the pump parameter $E$
for $\psi =0.3$, $\tilde{\gamma}_{0}^{\prime }=2$ and a 20\% mismatch ($%
\tilde{\delta}=0.1$). The resonance frequencies $\tilde{\omega}_{2}$ and $%
\tilde{\omega}_{4}$ start from the origin and grow approximately like $\sqrt{%
E}$. The damping parameter $w_{1}$ starts from zero and rapidly reaches the
constant value of $2$, whereas $w_{2}=2$ is independent of $E$, and $w_{4}$
moves quickly away from $2$ toward $1$. This behavior is not very sensitive
to the detuning angle.

At resonance $D^{\prime }$ factorizes into $i\gamma ^{\left( 3\right)
}D_{+}D_{-}$, being $D_{-}$, and $D_{+}$ two polynomials of second and third
order respectively, 
\[
D_{-}\left( \tilde{\omega}\right) =\tilde{\omega}^{2}-i\gamma ^{\left(
1\right) }\tilde{\omega}-2\gamma ^{\left( 2\right) }\;,\qquad D_{+}\left( 
\tilde{\omega}\right) =\tilde{\omega}^{3}-i\gamma ^{\left( 1\right) }\tilde{%
\omega}^{2}-\gamma ^{\left( 2\right) }\tilde{\omega}+i\gamma ^{\left(
3\right) } 
\]
where $\gamma ^{\left( 1\right) }=2+\tilde{\gamma}_{0}^{\prime }$, $\gamma
^{\left( 2\right) }=2E\tilde{\gamma}_{0}^{\prime }$, and $\gamma ^{\left(
3\right) }=4\left( E-1\right) \tilde{\gamma}_{0}^{\prime }\left( 1-\tilde{%
\delta}^{2}\right) $.

$D_{-}$ has two roots $\tilde{\Omega}_{4},\tilde{\Omega}_{5}$ which are
purely imaginary if $\gamma ^{\left( 1\right) 2}>8\gamma ^{\left( 2\right) }$%
, otherwise $\tilde{\Omega}_{4}=-\tilde{\Omega}_{5}^{\ast }.$ In turn $D_{+}$
has three purely imaginary roots ($\tilde{\Omega}_{1}=iw_{1}$, $\tilde{\Omega%
}_{2}=iw_{2}$, and $\tilde{\Omega}_{3}=iw_{3}$) if 
\[
\gamma ^{\left( 1\right) 2}>3\gamma ^{\left( 2\right) },\quad \gamma
^{\left( 1\right) }\gamma ^{\left( 2\right) }>9\gamma ^{\left( 3\right)
},\quad \left( \gamma ^{\left( 1\right) 2}-4\gamma ^{\left( 2\right)
}\right) \gamma ^{\left( 2\right) 2}>27\gamma ^{\left( 3\right) 2}+2\left(
2\gamma ^{\left( 1\right) 3}-9\gamma ^{\left( 1\right) }\gamma ^{\left(
2\right) }\right) \gamma ^{\left( 3\right) } 
\]
According to these inequalities (derived by means of the Sturm sequence) the
roots are all purely imaginary for $1<E<E_{\max }$, with $E_{\max }$
depending on $\tilde{\gamma}_{0}^{\prime }$ and $\tilde{\delta}$. In the
example plotted in Fig. 2 for $\tilde{\delta}=0.05$, $E_{\max }$ grows with $%
\tilde{\gamma}_{0}^{\prime }$ except for the presence of a peninsula in the
interval $\tilde{\gamma}_{0}^{\prime }\simeq 1.5\div 2.75$. Scaling the
threshold with $\tilde{\gamma}_{0}^{\prime }$ (Eq. (\ref{threshold})), for $%
\tilde{\gamma}_{0}^{\prime }\gg 1$ the roots are purely imaginary for all
realistic values of the excitation parameters. This explains why the
relaxation oscillations (resonances with non-vanishing real parts) can be
observed only for $\gamma _{0}^{\prime }$ not much greater than $\gamma $.

While the poles of $K_{jk\;}^{\frac{\pi }{2}}$ control the transient
behavior of $\left\langle \varphi _{j}^{\prime 2}\left( \tau ,0\right)
\right\rangle $, the relaxation oscillations correspond to those of $%
K_{jk}^{0}$. In particular, for $E$ rather large 
\begin{equation}
\tilde{\Omega}_{1}\simeq i\left( 1-\tilde{\delta}^{2}\right) \qquad \tilde{%
\Omega}_{2}\simeq \frac{i}{2}\left( 1+\tilde{\delta}^{2}+\tilde{\gamma}%
_{0}^{\prime }\right) {+}\sqrt{2E\tilde{\gamma}_{0}^{\prime }}
\label{relaxation}
\end{equation}
Similar expressions were obtained in Ref \cite{Lee} on the basis of a rate
equation model for the relaxation oscillations.

In Fig. 3 we have plotted the real and imaginary parts of $\tilde{\Omega}%
_{2,3}$ versus $1<E<10$ for different decay rates $\tilde{\gamma}%
_{0}^{\prime }=0.5,\,2.5,\,4.5,\,$and $6.5$ and $\delta =0$. These plots
also display the dependence of the roots on the detuning angle $\psi $ of a
balanced system having an effective excitation parameter $E_{eff}$ equal to $%
E$.

In the adiabatic limit ($\gamma _{0}^{\prime }\gg \gamma $ , see f.i. Ref. 
\cite{FabreI}) $D^{\prime }\left( \tilde{\omega}\right) $ reduces to a third
order polynomial, 
\[
D^{\prime }\left( \tilde{\omega}\right) =\tilde{\omega}^{3}+D^{\left(
3\right) }\tilde{\omega}^{2}+D^{\left( 2\right) }\tilde{\omega}+D^{\left(
1\right) } 
\]
that is the OPO is characterized only by four poles, $\tilde{\Omega}_{0}=0$, 
$\tilde{\Omega}_{1}=iw_{1}$, $\tilde{\Omega}_{2}=-\tilde{\Omega}_{3}^{*}=%
\tilde{\Omega}_{2}+iw_{2}$ and the relaxation oscillations disappear.

\section{\bf Transfer functions}

The $K_{kl}^{0,\frac{\pi }{2}}$ are singular in the half-plane $%
\mathop{\rm Im}%
\left( \tilde{\Omega}\right) \geq 0$, their poles coinciding with some
resonances $\tilde{\Omega}_{r}$ of the whole system. Their expressions for
the general case, together with those for the adiabatic case and for a
resonant device are reported in Appendix A.

For $\left| \omega \right| \rightarrow \infty $, $\;K_{jj\;}^{0,\frac{\pi }{2%
}}\rightarrow \frac{1}{\Delta _{j}}$, $K_{jj^{\prime }}^{0}$, $%
-K_{jj^{\prime }}^{\frac{\pi }{2}}\rightarrow \frac{1}{\Delta _{j}^{\ddagger
}\Delta _{j^{\prime }}}\;$and$\;\;K_{j0}^{0,\frac{\pi }{2}}\rightarrow \frac{%
1}{\Delta _{j}^{\ddagger }\Delta _{0}}$.

$K_{jk}^{0}$ is not singular at the origin contrariwise to $K_{jj}^{\frac{%
\pi }{2}}$ and $K_{jj^{\prime }}^{\frac{\pi }{2}}$. In particular in
proximity of the origin 
\begin{eqnarray}
K_{j0}^{0}\left( \omega \right) &\approx &\frac{1}{4}\left( 1-i\frac{\tan
\psi }{E_{eff}1}\right)  \nonumber \\
K_{jj}^{\frac{\pi }{2}}\left( \omega \right) &\approx &-K_{jj^{\prime }}^{%
\frac{\pi }{2}}\left( \omega \right) \approx i\frac{1-\tilde{\delta}^{2}}{%
4\cos ^{2}\psi }\frac{1}{\omega }  \nonumber \\
K_{j0}^{\frac{\pi }{2}}\left( \omega \right) &\approx &\frac{\left(
E_{eff}-1\right) -\left( \tan ^{2}\psi -\left( E_{eff}-1\right)
+iE_{eff}\tan \psi \right) \tilde{\delta}_{j}}{2E_{eff}}
\label{low-frequency}
\end{eqnarray}
where $\left( E_{eff}-1\right) =\left( {\cal E}-1\right) /\cos \psi $
represents the effective distance from the threshold of a detuned device.

Analogously the phase reduces for $\omega \approx 0$ to

\begin{equation}
\hat{\phi}_{\beta _{j}}\left( \omega \right) \approx -\frac{1-\tilde{\delta}%
^{2}}{4\cos ^{2}\psi \;}\frac{1}{\omega }\left( \hat{X}_{j}^{\prime 0}-\hat{X%
}_{j^{\prime }}^{\prime 0}\right) -\frac{\left( E_{eff}-1\right) -\left(
\tan ^{2}\psi -\left( E_{eff}-1\right) +iE_{eff}\tan \psi \right) \tilde{%
\delta}_{j}}{2E_{eff}}\otimes \hat{X}_{\phi _{p}}^{\frac{\pi }{2}+\psi _{p}}
\label{phase-origin}
\end{equation}
namely the detuning enhances, through the factors $\cos ^{-2}\psi $, the
phase fluctuations associated to the quantum noise sources.

The mathematical difficulties associated with the singularity of $K_{jj}^{%
\frac{\pi }{2}},K_{jj^{\prime }}^{\frac{\pi }{2}}$ at the origin can be
avoided by displacing slightly the zero $\Omega _{0}=0\mapsto \Omega
_{0}=i\varepsilon $ ($\varepsilon >0$), and letting $\varepsilon \rightarrow
0$ once performed the integrations.

The presence of poles with real parts larger than the imaginary ones is
evidenced in the plots of $\left| K_{jj}^{0}\right| ^{2}$, $\left|
K_{jj^{\prime }}^{0}\right| ^{2}$ and $\left| K_{j0}^{0}\right| ^{2}$
reported in Fig. 4 for cavity parameters $\tilde{\gamma}_{0}^{\prime }=1$,$\;%
\tilde{\delta}=0.05$, and $E=2,3,4$. They exhibit a behavior typical of
triply resonant devices, namely peaked at frequencies increasing with the
pump rate. While the peak height of the self--coupled response $K_{jj}^{0}$
(Figs 4a) slightly decreases with $E$, those of the cross--coupling $%
K_{jj^{\prime }}^{0}$ (Figs 4b) and of the response $K_{j0}^{0}$ (Figs 4c)
to pump fluctuations keep, respectively, nearly constant and growing in
height with $E$. All these frequency responses are not very sensitive to the
detuning.

\section{\bf Spectral densities}

For vacuum input fields, only antinormally ordered terms contribute to the
variances, i.e. 
\[
\left\langle \hat{b}_{in_{j}}\left( \omega \right) \hat{b}%
_{in_{j}}^{\ddagger }\left( \omega ^{\prime }\right) \right\rangle =\delta
\left( \omega ^{*}+\omega ^{\prime }\right) 
\]
Consequently, the noise correlations have in the frequency domain the
following expressions: 
\begin{eqnarray}
\left\langle \hat{Z}_{j}^{\prime }\left( \omega \right) \hat{Z}_{j}^{\prime
\ddagger }\left( \omega ^{\prime }\right) \right\rangle ^{\prime } &=&\frac{%
2\gamma _{j}^{\prime }}{\left| \kappa _{j}^{\prime }\right| }=\varsigma
_{j}^{\prime }  \nonumber \\
\left\langle \hat{Z}_{0}^{\prime }\left( \omega \right) \hat{Z}_{0}^{\prime
\ddagger }\left( \omega ^{\prime }\right) \right\rangle ^{\prime } &=&\frac{%
2\gamma _{0}^{\prime }}{\left| \kappa _{0}^{\prime }\right| }\frac{1}{{\cal E%
}-1}=\varsigma _{0}^{\prime }  \nonumber \\
\left\langle \hat{Z}_{\epsilon }\left( \omega \right) \hat{Z}_{\epsilon
}^{\ddagger }\left( \omega ^{\prime }\right) \right\rangle ^{\prime }
&=&\left\langle \hat{Z}_{\epsilon }^{\ddagger }\left( \omega \right) \hat{Z}%
_{\epsilon }\left( \omega \right) \right\rangle ^{\prime }=\left( \frac{E}{%
{\cal E}-1}\right) ^{2}S_{\epsilon }\left( \omega \right)  \nonumber \\
\left\langle \hat{Z}_{\phi }\left( \omega \right) \hat{Z}_{\phi }^{\ddagger
}\left( \omega ^{\prime }\right) \right\rangle ^{\prime } &=&\left\langle 
\hat{Z}_{\phi }^{\ddagger }\left( \omega \right) \hat{Z}_{\phi }\left(
\omega \right) \right\rangle ^{\prime }=\left( \frac{E}{{\cal E}-1}\right)
^{2}S_{\phi }\left( \omega \right)  \label{noise-correlations}
\end{eqnarray}
all the other terms vanishing identically. $S_{\epsilon }\left( \omega
\right) =\left\langle \left| \hat{\mu}_{\epsilon }\left( \omega \right)
\right| ^{2}\right\rangle ^{\prime }$ and $S_{\phi }\left( \omega \right)
=\Delta \nu _{L}^{2}/\omega ^{2}$ stand for the spectral densities of the
pump amplitude and phase respectively. The apex $(\left\langle
{}\right\rangle ^{\prime })$ indicates the omission on the right-side of the
factor $\delta \left( \omega ^{*}+\omega ^{\prime }\right) /C^{2}$.

\subsection{Gaussian statistics}

As a result of the superposition of uncorrelated spectral components, the
single beam fluctuation amplitude $\delta a\left( t\right) =k\tilde{a}$ is
proportional through a coefficient $k$ to the annihilation operator $\tilde{a%
}$ of an oscillator associated to a Gaussian density operator (see \cite
{Gardiner-Zoller,Hida}) 
\begin{equation}
\rho ={\cal N}e^{-\left( n_{\mu }+n_{\phi }\right) K_{0}+\frac{n_{\mu
}-n_{\phi }}{2}\left( K_{+}+K_{-}\right) }={\cal N}%
e^{2uK_{0}}e^{vK_{+}}e^{-wK_{-}}  \label{density-matrix}
\end{equation}
with $2K_{+}=\tilde{a}^{\dagger 2}\;,\;2K_{-}=\tilde{a}^{2}\;,\;4K_{0}=%
\tilde{a}^{\dagger }\tilde{a}+\tilde{a}\tilde{a}^{\dagger }$ forming a
realization of the SU(1,1) group, with the relevant algebraic structure
displayed by the commutators $\left[ K_{-},K_{+}\right] =2K_{0}\;,\;\left[
K_{0},K_{\pm }\right] =\pm K_{\pm }.$ In the above equation $\rho $ has been
expressed either as a single exponential depending on the coefficients $%
n_{\mu }$ and $n_{\phi }$ or as a disentangled product of exponentials
depending on $u,v$ and $w.$ The connection between these two representations
is displayed by the relations 
\[
2u=\ln \left( -w/v\right) \ ,\ v=\frac{\sinh \sqrt{n_{\mu }n_{\phi }}\sinh
\left( \sqrt{n_{\mu }n_{\phi }}+\Theta \right) }{\sinh ^{2}\Theta }\ ,\ w=-%
\frac{\sinh \sqrt{n_{\mu }n_{\phi }}}{\sinh \left( \sqrt{n_{\mu }n_{\phi }}%
+\Theta \right) } 
\]
with $\cosh \Theta =\left( n_{\mu }+n_{\phi }\right) /\left( n_{\mu
}-n_{\phi }\right) $. Here we limit ourselves to report the results of
calculations based on the Wei-Norman technique \cite{Wei} (see also \cite
{Dattoli}) which will be discussed elsewhere together with the calculation
of the trace 
\[
Tr\left( e^{2uK_{0}}e^{vK_{+}}e^{-wK_{-}}\right) =\frac{1}{\sqrt{\left(
vw+1\right) e^{u}-2+e^{-u}}} 
\]

The moments of $K_{0}$ and $K_{\pm }$ can be easily obtained by
differentiating the operator $e^{2uK_{0}}e^{vK_{+}}e^{-wK_{-}}$ with respect
to $u$, $v$ and $w$. In particular,

\[
\left\langle \left[ \phi ,\mu \right] \right\rangle =i\frac{k^{2}}{2}\ ,\
\left\langle \mu ^{2}\right\rangle =\frac{k^{2}}{2}\left( \frac{1}{n_{\mu }}+%
\frac{1}{2}\right) ,\ \left\langle \phi ^{2}\right\rangle =\frac{k^{2}}{2}%
\left( \frac{1}{n_{\phi }}+\frac{1}{2}\right) 
\]

It is straightforward to extend the above representation to the twin pair $%
\delta a_{1}\left( t\right) =k\tilde{a}$, $\delta a_{2}\left( t\right) =k%
\tilde{b}$ for a balanced OPO, 
\[
\rho ={\cal N}\exp \left( C_{0}^{+}K_{0}^{+}+C_{0}^{-}K_{0}^{-}+C^{+}\left(
K_{+}^{+}+K_{-}^{+}\right) +C^{-}\left( K_{+}^{-}+K_{-}^{-}\right) \right) 
\]
with $4K_{\pm }^{+}=\left\{ \tilde{a}^{\dagger }/\tilde{a}\right\}
^{2}+\left\{ \tilde{b}^{\dagger }/\tilde{b}\right\} ^{2}$, $%
2K_{0}^{+}=K_{0a}+K_{0b}$, $2K_{\pm }^{-}=\left\{ \tilde{a}\tilde{b}/\tilde{a%
}^{\dagger }\tilde{b}^{\dagger }\right\} \;$and$\;4K_{0}^{-}=\tilde{a}%
^{\dagger }\tilde{b}+\tilde{a}\tilde{b}^{\dagger }$ forming an enveloping
algebra of SU(1,1) characterized by the algebraic structure $2\left[
K_{0}^{\pm },K_{\pm }^{\pm /\mp }\right] =\pm K_{\pm }^{+/-},2\left[
K_{-}^{\pm },K_{+}^{\mp }\right] =K_{0}^{-}$ all the other commutators
vanishing identically. The disentangling of this operator will be discussed
elsewhere.

Using the squeezing operator $S\left( \theta \right) =\exp \left[ \frac{%
\theta }{2}\left( \tilde{a}^{2}-\tilde{a}^{\dagger 2}\right) \right] $ with $%
\tanh \left( 2\theta \right) =1/\cosh \Theta $, introduced by Gardiner and
Zoller \cite{Gardiner-Zoller} for transforming a gaussian density matrix
into a thermal oscillator, we obtain in the present case $\rho =2\sinh \frac{%
\sqrt{n_{\mu }n_{\phi }}}{2}\exp \left( -2\sqrt{n_{\mu }n_{\phi }}%
K_{0}\right) $. Accordingly the twin beam fluctuations can be represented as
a pair of squeezed and coupled thermal oscillators.

Finally, from the Gaussian statistics it descends for the phase factors $%
e^{i\phi _{j}\left( \tau \right) }$ correlation 
\begin{equation}
\left\langle e^{i\phi _{j}\left( \tau \right) }e^{-i\phi _{j}\left( 0\right)
}\right\rangle =e^{\left\langle \left( \phi _{j}\left( \tau \right) -\phi
_{j}\left( 0\right) \right) \phi _{j}\left( 0\right) \right\rangle }
\label{phase-factor-correlations}
\end{equation}

\subsection{Correlations and frequency spectra}

The various correlations can be represented as combinations of the functions 
\begin{eqnarray}
K_{\phi _{j}\phi _{l}}\left( \tau \right) &=&\sum_{k}\zeta _{k}^{\prime
}\sum_{r=1}^{5}e^{i\Omega _{r}\tau }K_{jk,r}^{\frac{\pi }{2}}K_{lk}^{\frac{%
\pi }{2}*}\left( \Omega _{r}^{*}\right)  \nonumber \\
K_{\mu _{j}\mu _{l}}\left( \tau \right) &=&\sum_{k}\zeta _{k}^{\prime
}\sum_{r=1}^{5}e^{i\Omega _{r}\tau }K_{jk,r}^{0}K_{lk}^{0*}\left( \Omega
_{r}^{*}\right)  \nonumber \\
K_{\mu _{j}\phi _{l}}\left( \tau \right) &=&\sum_{k}\zeta _{k}^{\prime
}\sum_{r=1}^{5}e^{i\Omega _{r}\tau }K_{jk,r}^{0}K_{lk}^{\frac{\pi }{2}%
*}\left( \Omega _{r}^{*}\right)  \nonumber \\
K_{\phi _{j}\mu _{l}}\left( \tau \right) &=&\sum_{k}\zeta _{k}^{\prime
}\sum_{r=1}^{5}e^{i\Omega _{r}\tau }K_{jk,r}^{\frac{\pi }{2}%
}K_{lk}^{0*}\left( \Omega _{r}^{*}\right)  \label{funzioni-K}
\end{eqnarray}
with $l=0,1,2$ , the sums being extended to the 5 roots $\Omega _{r}$ of the
characteristic polynomial $D^{\prime }$ (i.e. except for $\Omega _{0}$), and 
$K_{jk,r}^{0,\frac{\pi }{2}}$ standing for the residues of $K_{jk}^{0,\frac{%
\pi }{2}}\left( \omega \right) $ at $\Omega _{r}$.

We have for the correlations 
\begin{eqnarray}
\left\langle \left( \phi _{j}\left( \tau \right) -\phi _{j}\left( 0\right)
\right) \phi _{j}\left( 0\right) \right\rangle &=&\frac{i}{4C^{2}}\left(
K_{\phi _{j}\phi _{j}}\left( \tau \right) -K_{\phi _{j}\phi _{j}}\left(
0\right) \right) -\Delta \nu _{j}\tau  \nonumber \\
\left\langle \mu _{j}\left( \tau \right) \mu _{j}\left( 0\right)
\right\rangle &=&\frac{i}{4C^{2}}K_{\mu _{j}\mu _{j}}\left( \tau \right) 
\nonumber \\
\left\langle \mu _{j}\left( \tau \right) \phi _{l}\left( 0\right)
\right\rangle -\left\langle \phi _{j}\left( \tau \right) \mu _{l}\left(
0\right) \right\rangle &=&\frac{1}{4C^{2}}\left( K_{\mu _{j}\phi _{l}}\left(
\tau \right) +K_{\phi _{j}\mu _{l}}\left( \tau \right) \right)
\label{correlations}
\end{eqnarray}
and the commutators 
\begin{eqnarray}
\left\langle \left[ \phi _{j}\left( 0\right) ,\phi _{j}\left( \tau \right) %
\right] \right\rangle &=&\frac{i}{2C^{2}}%
\mathop{\rm Re}%
\left\{ K_{\phi _{j}\phi _{j}}\left( \tau \right) \right\}  \nonumber \\
\left\langle \left[ \mu _{j}\left( 0\right) ,\mu _{j}\left( \tau \right) %
\right] \right\rangle &=&-\frac{i}{2C^{2}}%
\mathop{\rm Re}%
\left\{ K_{\mu _{j}\mu _{j}}\left( \tau \right) \right\}  \nonumber \\
\left\langle \left[ \phi _{j}\left( 0\right) ,\mu _{j}\left( \tau \right) %
\right] _{+}\right\rangle &=&\frac{1}{2C^{2}}%
\mathop{\rm Im}%
\left\{ K_{\phi _{j}\mu _{j}}\left( \tau \right) \right\}  \nonumber \\
\left\langle \left[ \phi _{j}\left( 0\right) ,\mu _{j}\left( \tau \right) %
\right] \right\rangle &=&\frac{i}{2C^{2}}%
\mathop{\rm Re}%
\left\{ K_{\phi _{j}\mu _{j}}\left( \tau \right) \right\}
\label{commutators}
\end{eqnarray}
with $\tau >0.$ While (\ref{commutators}-a,b,c) vanish at resonance, the
last expression does not. For a detuned OPO $\mu _{j}\left( t\right) $ and $%
\phi _{j}\left( t\right) $ do not commute with the respective quantities at
different times while $\mu _{j}\left( t\right) $ does not anticommute with $%
\phi _{j}\left( t^{\prime }\right) $. By similar reasoning it can be also
shown $\left\langle \phi _{j}\left( 0\right) \dot{\phi}_{j}\left( 0\right)
\right\rangle \neq 0$, that is the instantaneous frequency is correlated
with the phase. In Fig. 5 the phase commutator has been plotted versus $\tau 
$ for an OPO with $\tilde{\gamma}_{0}^{\prime }=4$, $\tilde{\delta}=0.2$ and 
$\psi =0.1$ and ${\cal E}=1.5$, $4$ (Eq.(\ref{Epsilon})). These plots
exhibit respectively oscillatory and damped behaviors since for ${\cal E}=4$
the roots have non vanishing real parts, while in the other case they are
almost purely imaginary.

The variance of the phase delay is given by 
\[
\left\langle \left( \phi _{j}\left( \tau \right) -\phi _{j}\left( 0\right)
\right) ^{2}\right\rangle =\frac{1}{2C^{2}}%
\mathop{\rm Im}%
\left\{ K_{\phi _{j}\phi _{j}}\left( \tau \right) -K_{\phi _{j}\phi
_{j}}\left( 0\right) \right\} +2\Delta \nu _{j}\tau 
\]
with 
\[
\Delta \nu _{j}=F_{\phi }\Delta \nu _{L}+\frac{1}{8C^{2}}\frac{\left( 1-%
\tilde{\delta}^{2}\right) ^{2}}{\cos ^{4}\psi } 
\]
$F_{\phi }$ standing for the broadening factor of the pump linewidth $\Delta
\nu _{L}$ (see below Eq. (\ref{Langevin})), 
\[
F_{\phi }=\left| \frac{\left( E_{eff}-1\right) -\left( \tan ^{2}\psi -\left(
E_{eff}-1\right) +iE_{eff}\tan \psi \right) \tilde{\delta}_{j}}{2E_{eff}}%
\frac{E\cos \psi }{E_{eff}-1}\right| ^{2} 
\]

In a similar fashion we have for the phase sum variance 
\[
\left\langle \left( \phi _{j}\left( \tau \right) +\phi _{j}\left( 0\right)
\right) ^{2}\right\rangle =\frac{1}{4C^{2}}%
\mathop{\rm Im}%
\left\{ K_{\phi _{j}\phi _{j^{\prime }}}\left( \tau \right) -K_{\phi
_{j^{\prime }}\phi _{j}}^{\ast }\left( \tau \right) +K_{\phi _{j}\phi
_{j}}\left( 0\right) -K_{\phi _{j^{\prime }}\phi _{j^{\prime }}}^{\ast
}\left( 0\right) \right\} +2\Delta \nu _{j}\tau 
\]
so that (see \cite{Reid}) in view of Eq. (\ref{phase-factor-correlations}) 
\begin{eqnarray*}
\left\langle e^{i\phi _{j}\left( \tau \right) }e^{-i\phi _{j}\left( 0\right)
}\right\rangle &=&e^{-\Delta \nu _{j}\tau }\left( 1+\frac{i}{4C^{2}}\left(
K_{\phi _{j}\phi _{j}}\left( \tau \right) -K_{\phi _{j}\phi _{j}}\left(
0\right) \right) \right) \\
\left\langle e^{-i\phi _{j}\left( \tau \right) }e^{-i\phi _{j^{\prime
}}\left( 0\right) }\right\rangle &=&e^{-\Delta \nu _{j}\tau }\left( 1+\frac{i%
}{4C^{2}}\left( K_{\phi _{j}\phi _{j^{\prime }}}\left( \tau \right) -K_{\phi
_{j}\phi _{j^{\prime }}}\left( 0\right) \right) \right)
\end{eqnarray*}
while the second-order correlations for the mode amplitudes $a_{j}$, $%
a_{j}^{\dagger }$ read 
\begin{eqnarray*}
\left\langle a_{j}^{\dagger }\left( \tau \right) a_{j}\left( 0\right)
\right\rangle &=&r_{j}^{2}e^{-\Delta \nu _{j}\tau }\left[ 1+\frac{i}{4C^{2}}%
\left( K_{\mu _{j}\mu _{j}}\left( \tau \right) +K_{\phi _{j}\phi _{j}}\left(
\tau \right) -K_{\phi _{j}\phi _{j}}\left( 0\right) -K_{\mu _{j}\phi
_{l}}\left( \tau \right) -K_{\phi _{j}\mu _{l}}\left( \tau \right) \right) %
\right] \\
\left\langle a_{j}\left( \tau \right) a_{j^{\prime }}\left( 0\right)
\right\rangle &=&r_{j}r_{j^{\prime }}e^{-\Delta \nu _{j}\tau }\left[ 1+\frac{%
i}{4C^{2}}\left( K_{\mu _{j}\mu _{j^{\prime }}}\left( \tau \right) +K_{\phi
_{j}\phi _{j^{\prime }}}\left( \tau \right) -K_{\phi _{j}\phi _{j^{\prime
}}}\left( 0\right) -K_{\mu _{j}\phi _{j^{\prime }}}\left( \tau \right)
-K_{\phi _{l}\mu _{l^{\prime }}}\left( \tau \right) \right) \right]
\end{eqnarray*}

Further, the intensity variances of the single beams are represented by 
\begin{equation}
\left\langle :a_{j}^{\dagger }\left( \tau \right) a_{j}\left( \tau \right)
,a_{j}^{\dagger }\left( 0\right) a_{j}\left( 0\right) :\right\rangle
=r_{j}^{4}\left( 2\left\langle \mu _{j}\left( \tau \right) \mu _{j}\left(
0\right) +\mu _{j}\left( 0\right) \mu _{j}\left( \tau \right) \right\rangle
+i\left\langle \left[ \phi _{j}\left( 0\right) ,\mu _{j}\left( \tau \right) %
\right] \right\rangle \right)  \label{intensity-variance}
\end{equation}
$:\cdot :$ standing for time-normal ordering. Analogously, for the
respective spectra, normalized with respect to $r_{j}^{4}/2C^{2},$%
\[
:S_{j}:=S_{j}+S_{\left[ \mu ,\phi \right] _{j}} 
\]
with $S_{j}$ separating in three terms 
\begin{equation}
S_{j}=S_{\mu _{j}}+\left( \frac{E}{{\cal E}-1}\right) ^{2}S_{\epsilon
}S_{\epsilon _{j}}+\left( \frac{E}{{\cal E}-1}\right) ^{2}S_{\phi }S_{\phi
_{j}}  \label{single-beam}
\end{equation}
representing in the order the quantum and the pump amplitude and phase
contributions 
\begin{eqnarray}
S_{\mu _{j}}\left( \omega \right) &=&\sigma _{\mu _{j}}^{2}\left( \omega
\right) +\sigma _{\mu _{j}}^{2}\left( -\omega \right)  \nonumber \\
S_{\epsilon _{j}}\left( \omega \right) &=&\sigma _{\epsilon _{j}}^{2}\left(
\omega \right) +\sigma _{\epsilon _{j}}^{2}\left( -\omega \right)  \nonumber
\\
S_{\phi _{j}}\left( \omega \right) &=&\sigma _{\phi _{j}}^{2}\left( \omega
\right) +\sigma _{\phi _{j}}^{2}\left( -\omega \right)
\label{noise-contributions}
\end{eqnarray}
$S_{\epsilon }$ and $S_{\phi }$ represent the pump amplitude and phase
spectral densities (see Eqs.(\ref{noise-correlations}--c, d)) while $\sigma
_{\mu _{j}}^{2}$,$\sigma _{\epsilon _{j}}^{2}$ and $\sigma _{\phi _{j}}^{2}$
stand for 
\begin{eqnarray}
\sigma _{\mu _{j}}^{2} &=&\sum_{k}\left| K_{jk}^{0}\right| ^{2}\varsigma
_{k}^{\prime }  \nonumber \\
\sigma _{\epsilon _{j}}^{2} &=&\left| \frac{e^{i\psi
_{p}}K_{j0}^{0}+e^{-i\psi _{p}}K_{j0}^{0\ddagger }}{2}\right| ^{2}  \nonumber
\\
\sigma _{\phi _{j}}^{2} &=&\left| \frac{e^{i\psi _{p}}K_{j0}^{0}-e^{-i\psi
_{p}}K_{j0}^{0\ddagger }}{2}\right| ^{2}  \label{contributions}
\end{eqnarray}

In turn, the normalized spectrum of the commutator $\left[ \mu _{j},\phi _{j}%
\right] $ splits in the components 
\[
S_{\left[ \mu ,\phi \right] _{j}}\left( \omega \right) =\sigma _{\left[ \mu
,\phi \right] _{j}}^{2}\left( \omega \right) +\sigma _{\left[ \mu ,\phi %
\right] _{j}}^{2}\left( -\omega \right) 
\]
with 
\begin{equation}
\sigma _{\left[ \mu ,\phi \right] _{j}}^{2}\left( \omega \right) =%
\mathop{\rm Re}%
\left\{ \sum_{k,r}\zeta _{k}^{\prime }\frac{K_{jk,r}^{0}}{\omega -\Omega _{r}%
}K_{jk}^{\frac{\pi }{2}\ast }\left( \Omega _{r}^{\ast }\right) \right\}
\label{spectral-density-commutator}
\end{equation}

Similar expressions hold for the output beams with $\bar{I}_{j}=2\gamma
_{j}r_{j}^{2}$ in place of $r_{j}^{2}$ and $\sigma _{\mu _{j}}^{2}$ (Eq. (%
\ref{contributions}--a)) replaced by 
\begin{equation}
\sigma _{\mu j}^{2}=\sum_{k\neq j}\left| K_{jk}^{0}\right| ^{2}\varsigma
_{k}^{\prime }\ +\left( \left| K_{jj}^{0}-\frac{\kappa _{j}^{\prime }}{%
2\gamma _{j}}\right| ^{2}\frac{\gamma _{j}}{\gamma _{j}^{\prime }}+\left|
K_{jj}^{0}\right| ^{2}\frac{\varkappa _{j}}{\gamma _{j}^{\prime }}\right)
\varsigma _{j}^{\prime }\   \label{external}
\end{equation}

In Fig. 6, we have plotted $S_{j}$ (Eq. (\ref{single-beam})), the spectrum
measured by analyzing the photocurrent with a spectrum analyzer, for
different detunings ($\psi =0$,\thinspace $0.25$,\thinspace $0.5$) and pump
parameter $E=3$ by omitting the laser excess noise and phase fluctuations.
For both $\tilde{\gamma}_{0}^{\prime }=4$ (top) and $0.5$ (bottom), in the
case of zero and small detuning, $S_{j}$ starts from below the SNL, while,
for $\psi =0.5$, it starts just above the SNL. In all cases $S_{j}$ rapidly
increases by reaching a peak in correspondence of the relaxation
oscillations frequency, and then decays monotonically toward the SNL. The
height of the peak increases notably either with the detuning or by reducing
the pump mode bandwidth.

Figure 7 describes the effects on the single beam spectrum of the pump
excess noise at different detuning angles. $\left( \frac{E}{{\cal E}-1}%
\right) ^{2}S_{\epsilon _{j}}\left( \omega \right) $ has been plotted for
different detunings ($\psi =0$,\thinspace $0.25$,\thinspace $0.5$), $\tilde{%
\delta}=0.05$, pump parameter $E=3$, and crystal losses different for the
signal/idler and the pump $\varkappa _{1,2}/\gamma =0.3$, $\varkappa
_{0}=\varkappa _{1}/3$. For both cavity losses $\tilde{\gamma}_{0}^{\prime
}=4$ (top) and $0.5$ (bottom ), $\left( \frac{E}{{\cal E}-1}\right)
^{2}S_{\epsilon _{j}}$ increases with $\psi $ and $\omega $ by reaching a
peak in correspondence of the relaxation oscillation. The behavior observed
for $0.5<\tilde{\gamma}_{0}^{\prime }<4$ leads us to conclude that for mode
dampings of the same order, the detuning confines the observable squeezing
to a small fraction of the twin--beam bandwidth. Moreover, it increases the
amplitude of the excess laser noise transferred to the single beam spectrum,
thus preventing the reduction below the shot noise level.

The difference spectrum $S_{d}$ is obtained from $S_{j}$ by replacing in (%
\ref{contributions}), (\ref{spectral-density-commutator}) and (\ref{external}%
) $K_{jk}^{0,\frac{\pi }{2}}$ with $\Delta K_{k}^{0,\frac{\pi }{2}}=$ $%
\left( \bar{I}_{1}K_{1k}^{0,\frac{\pi }{2}}-\bar{I}_{2}K_{2k}^{0,\frac{\pi }{%
2}}\right) /\left( \bar{I}_{1}+\bar{I}_{2}\right) $ and analogously for the
residues, $\frac{\kappa _{j}^{\prime }}{2\gamma _{j}}$ and $\frac{\gamma _{j}%
}{\gamma _{j}^{\prime }}$. In particular, the contributions of pump
amplitude and phase fluctuations are represented by Eqs.(\ref
{noise-contributions}--b,c) respectively, with $\sigma _{\epsilon _{j}}^{2}$
and $\sigma _{\phi _{j}}^{2}$ replaced by $\sigma _{\epsilon _{d}}^{2}$ and $%
\sigma _{\phi _{d}}^{2}$, which are in turn defined by equations similar to (%
\ref{contributions}--b, c) with $K_{j0}^{0}$ replaced by $\Delta
K_{0}^{0}=\left( \bar{I}_{1}K_{10}^{0}-\bar{I}_{2}K_{20}^{0}\right) /\left( 
\bar{I}_{1}+\bar{I}_{2}\right) .$ Being the spectral contributions of the
laser excess noise and phase diffusion concentrated in the low frequency
region, it is worth noting that (see Eq. (\ref{low-frequency}-a)) 
\[
\Delta K_{0}^{0}\left( 0\right) =\left( 1-i\frac{\tan \psi }{E_{eff}}\right) 
\frac{\tilde{\delta}\varkappa }{1-\tilde{\delta}^{2}}\cos \psi 
\]
so that

\begin{eqnarray}
\sigma _{\epsilon _{d}}^{2}\left( 0\right) &=&\left( \frac{\tilde{\delta}%
\varkappa }{1-\tilde{\delta}^{2}}\right) ^{2}\left| \cos \psi \cos \psi _{p}+%
\frac{\sin \psi }{E_{eff}}\sin \psi _{p}\right| ^{2}  \nonumber \\
\sigma _{\phi _{d}}^{2}\left( 0\right) &=&\left( \frac{\tilde{\delta}%
\varkappa }{1-\tilde{\delta}^{2}}\right) ^{2}\left| \cos \psi \sin \psi _{p}-%
\frac{\sin \psi }{E_{eff}}\cos \psi _{p}\right| ^{2}
\label{difference-noise}
\end{eqnarray}
These expressions evidence the deleterious effects of crystal losses and
detuning also in the difference spectra. While for a resonant device only
the excess noise influences the spectrum, in a detuned device also the phase
diffusion contributes to the noise in proximity of the origin.

The plots of Fig. 8 show the difference spectrum $S_{d}$ for an unbalanced
resonant OPO ($\tilde{\delta}=0.05$, $\psi =0$) for $\tilde{\gamma}%
_{0}^{\prime }=4$ (top) , $0.5$ (bottom). The peak observed in the single
beam spectrum for $\tilde{\gamma}_{0}^{\prime }=0.5$ (Figs. 6 and 7--bottom)
is present also in the difference spectrum for the same damping ratio
whereas for $\tilde{\gamma}_{0}^{\prime }=4$ (top) the difference spectrum
keeps unchanged its sub--shot--noise character.

\section{\bf Conclusions}

The steady-state characteristics of the beams generated by an OPO depend on
the decay constants of the cavity modes, the non linear crystal losses, the
excitation parameter $E$, and the deviation from resonance (detuning). The
last condition has been represented by an angle $\psi $. While for resonant
devices the beam amplitudes are proportional to $\sqrt{E-1}$, in presence of
detuning $E$ has been replaced by an effective excitation parameter ${\cal E}
$. Similarly, the phase difference between the driving field and the pump
mode is a function of the detuning angle $\psi $. The effects of these
changes propagate to the fluctuating parts of the twin beams amplitudes.

Once linearized, the fluctuations of the twin beams are characterized by
resonance frequencies which control the frequency responses $K_{kl}^{0,\frac{%
\pi }{2}}$ to the different noise sources (quantum noise of
signal/idler/pump and phase-amplitude fluctuations of the pump laser). These
functions have in general six poles $\Omega _{r}$ in the half plane $%
\mathop{\rm Im}%
\left( \omega \right) \geq 0$. One, $\Omega _{0}=0$, always lies at the
origin, a second one, $\Omega _{1}=iw_{1}$, is located on the positive
imaginary axis, while the remaining four ones may be purely imaginary or
complex; when complex ($\Omega _{2}=-\Omega _{3}^{*}$, $\Omega _{4}=-\Omega
_{5}^{*}$) they are characterized by two damping constants $w_{2},w_{4}$ and
frequencies $\omega _{2},\omega _{4}$.

At resonance the $K_{kl}^{0}$ have only three poles, different from the
those of $K_{kl}^{\frac{\pi }{2}}$. Their locations in the complex plane
depend on the excitation parameter $E$, the cavity damping coefficient $%
\tilde{\gamma}_{0}^{\prime }$ at the pump frequency, and the mismatch
parameter $\tilde{\delta}$. For given $\tilde{\delta}$ the couple of
parameters $\left( E,\tilde{\gamma}_{0}^{\prime }\right) $ corresponding to
purely imaginary resonances forms, on the $E$--$\tilde{\gamma}_{0}^{\prime }$
plane, a connected region delimited by the straight line $E=1$ and a curve
representing the maximum excitation $E_{\max }$ for which the poles are all
imaginary. For $\tilde{\gamma}_{0}^{\prime }$ comprised in a particular
interval the poles are imaginary for $1<E<E_{1}$ and $E_{2}<E<E_{\max }$.
Outside this region the two poles are complex conjugate and the spectrum
exhibits a relaxation oscillation peak \cite{Lee,Porzio}.

The amplitude quadrature responses $K_{jk}^{0}$ to the various noise terms
are regular for $\omega \rightarrow 0$ , contrariwise to the phase
quadratures, $K_{jj}^{\frac{\pi }{2}}$ and $K_{jj^{\prime }}^{\frac{\pi }{2}%
} $ which present a pole. The difficulties of this pole have been bypassed
by displacing it from the origin by a small quantity $\varepsilon $, and
letting $\varepsilon \rightarrow 0$ at the end of the calculations. So doing
the complex fluctuation $\mu -i\phi $ of each mode emerges as proportional
to the annihilation operator $\tilde{a}$ of a squeezed thermal oscillator $%
\left( \mu -i\phi =k\tilde{a}\right) $.

By exploiting the Gaussian statistics of the fluctuations second and fourth
order correlations have been expressed by suitable combinations of the
frequency responses and their residues. So doing, it has been filled the gap
between the evaluation of the variance and covariance of the fields carried
out by Reid and Drummond \cite{Reid} by means of classical stochastic
equations, limited to resonant and symmetric OPO, and that based on quantum
Langevin equations. The operator representation offers the opportunity of
using the commutators of the field variables at different times for
characterizing the respective correlations and gaining further insight in
the detuning effects. In fact, the complex damping coefficients make
asymmetric the frequency responses in presence of detuning, $\left|
K_{kl}^{0,\frac{\pi }{2}}\left( \omega \right) \right| ^{2}\neq \left|
K_{kl}^{0,\frac{\pi }{2}}\left( -\omega \right) \right| ^{2}$, thus implying
that $\mu _{j}\left( t\right) $ and $\phi _{j}\left( t\right) $ at different
times do not commute as does the instantaneous frequency with the phase at
the same time $\left( \left\langle \phi _{j}\left( 0\right) \dot{\phi}%
_{j}\left( 0\right) \right\rangle \neq 0\right) $.

The detuning interplays with pump excess noise and phase diffusion by
increasing the spectral broadening of the twin beams and enhancing the low
frequency part of the spectra. These negative effects are mitigated by a
well balanced configuration of the OPO.

While triply resonant devices reduce the threshold levels by allowing the
investigation of OPO quantum properties for large excitation parameters,
their spectra differ notably from those typical of the adiabatic regime. The
single beam spectral density starts, for $\omega =0$, from below the SNL,
rapidly increases by reaching a peak in correspondence of the relaxation
frequency, and then decays monotonically toward the SNL. The height of the
peak increases notably as the pump decay constant decreases with respect to
those of the signal/idler. These oscillations tend to obscure the expected
single beam squeezing, by confining it to a low frequency region where the
fluctuation spectrum is mostly conditioned by the laser excess noise and
phase diffusion.

\section{Appendix A}

$A_{k}\left( \omega \right) $ and $B_{k}\left( \omega \right) $ depend on
the functions $\Delta _{k}\left( \omega \right) $ 
\begin{eqnarray*}
A_{j}=\Delta _{j}+\frac{\Delta _{j^{\prime }}}{\Delta _{j^{\prime }}\Delta
_{0}+1}-\frac{\Delta _{0}^{\ddagger }}{\Delta _{j^{\prime }}^{\ddagger
}\Delta _{0}^{\ddagger }+1}\;,\qquad &&A_{0}=\Delta _{0}+\frac{\Delta
_{1}^{\ddagger }}{\Delta _{2}\Delta _{1}^{\ddagger }-1}+\frac{\Delta
_{2}^{\ddagger }}{\Delta _{1}\Delta _{2}^{\ddagger }-1} \\
B_{j}=\frac{1}{\Delta _{j^{\prime }}\Delta _{0}+1}+\frac{1}{\Delta
_{j^{\prime }}^{\ddagger }\Delta _{0}^{\ddagger }+1}\;,\qquad &&B_{0}=\frac{1%
}{\Delta _{2}\Delta _{1}^{\ddagger }-1}+\frac{1}{\Delta _{1}\Delta
_{2}^{\ddagger }-1}
\end{eqnarray*}
while the $C_{k}\left( \omega \right) $ are linear combinations of the noise
Fourier transforms $\tilde{Z}_{k}^{\prime }$ (see Eqs. $\left( \ref
{TRO-final}\right) $)

\[
C_{j}=\tilde{Z}_{j}^{\prime }+\frac{\Delta _{j}\tilde{Z}_{0}^{\prime }-%
\tilde{Z}_{j^{\prime }}^{\prime }}{\Delta _{j^{\prime }}\Delta _{0}+1}+\frac{%
\tilde{Z}_{0}^{\prime \ddagger }+\Delta _{0}^{\ddagger }\tilde{Z}_{j^{\prime
}}^{\prime \ddagger }}{\Delta _{j^{\prime }}^{\dagger }\Delta _{0}^{\dagger
}+1}\;,\qquad C_{0}=\tilde{Z}_{0}^{\prime }-\frac{\Delta _{1}^{\ddagger }%
\tilde{Z}_{2}^{\prime }+\tilde{Z}_{1}^{\prime \ddagger }}{\Delta _{2}\Delta
_{1}^{\ddagger }-1}-\frac{\Delta _{2}^{\ddagger }\tilde{Z}_{1}^{\prime }+%
\tilde{Z}_{2}^{\prime \ddagger }}{\Delta _{1}\Delta _{2}^{\ddagger }-1} 
\]

For a detuned system ($\psi \neq 0$) $A_{k}$ differs from the adjoint $%
A_{k}^{\ddagger }$, while $B_{k}$ coincides with $B_{k}^{\ddagger }$.

Next, introducing the functions 
\[
F_{k\pm }=\frac{A_{k}\pm B_{k}}{A_{k}A_{k}^{\ddagger }-B_{k}^{2}} 
\]
the frequency responses $K_{kl}^{0,\frac{\pi }{2}}$ defined by Eqs. (\ref
{quadratures}) are given by 
\begin{eqnarray*}
K_{jj}^{0}=F_{j-}^{\ddagger }\;,\qquad &&K_{jj^{\prime }}^{0}=\frac{%
F_{j-}\Delta _{0}-F_{j-}^{\ddagger }}{\Delta _{j^{\prime }}\Delta _{0}+1} \\
K_{j0}^{0}=\frac{F_{j-}^{\ddagger }\Delta _{j^{\prime }}+F_{j-}}{\Delta
_{j^{\prime }}\Delta _{0}+1}\;,\qquad &&K_{0j}^{0}=-\frac{%
F_{0-}+F_{0-}^{\ddagger }\Delta _{j^{\prime }}^{\ddagger }}{\Delta
_{j}\Delta _{j^{\prime }}^{\ddagger }-1}
\end{eqnarray*}
and 
\[
K_{jj}^{\frac{\pi }{2}}=F_{j+}^{\ddagger }\;,\qquad K_{jj^{\prime }}^{\frac{%
\pi }{2}}=-\frac{F_{j+}\Delta _{0}+F_{j+}^{\ddagger }}{\Delta _{j^{\prime
}}\Delta _{0}+1}\;,\qquad K_{j0}^{\frac{\pi }{2}}=\frac{F_{j+}^{\ddagger
}\Delta _{j^{\prime }}-F_{j+}}{\Delta _{j^{\prime }}\Delta _{0}+1} 
\]

Expressing $F_{k\pm }$ and $\Delta _{k}$ as functions of $\tilde{\omega}$
yields

\begin{eqnarray}
K_{jj}^{0} &=&-i\frac{\left( 1+\tilde{\delta}_{j}\right) e^{i\psi }}{\cos
\psi }\frac{\tilde{\omega}^{4}+K_{jj}^{0\left( 5\right) }\tilde{\omega}%
^{3}+K_{jj}^{0\left( 4\right) }\tilde{\omega}^{2}+K_{jj}^{0\left( 3\right) }%
\tilde{\omega}+K_{jj}^{0\left( 2\right) }}{D^{\prime }}  \nonumber \\
K_{jj^{\prime }}^{0} &=&-\frac{1-\tilde{\delta}^{2}}{\cos ^{2}\psi }\frac{%
\tilde{\omega}^{3}+K_{jj^{\prime }}^{0\left( 4\right) }\tilde{\omega}%
^{2}+K_{jj^{\prime }}^{0\left( 3\right) }\tilde{\omega}+K_{jj^{\prime
}}^{0\left( 2\right) }}{D^{\prime }}  \nonumber \\
K_{j0}^{0} &=&\tilde{\gamma}_{0}^{\prime }\left( E_{eff}-1\right) \left( 1+%
\tilde{\delta}_{j}\right) \frac{\tilde{\omega}^{3}+K_{j0}^{0\left( 4\right) }%
\tilde{\omega}^{2}+K_{j0}^{0\left( 3\right) }\tilde{\omega}+K_{j0}^{0\left(
2\right) }}{D^{\prime }}  \label{Kappas0}
\end{eqnarray}
and 
\begin{eqnarray}
K_{jj}^{\frac{\pi }{2}} &=&i\frac{\left( 1+\tilde{\delta}_{j}\right)
e^{i\psi }}{\cos \psi }\frac{\tilde{\omega}^{5}+K_{jj}^{\frac{\pi }{2}\left(
5\right) }\tilde{\omega}^{4}+K_{jj}^{\frac{\pi }{2}\left( 4\right) }\tilde{%
\omega}^{3}+K_{jj}^{\frac{\pi }{2}\left( 3\right) }\tilde{\omega}%
^{2}+K_{jj}^{\frac{\pi }{2}\left( 2\right) }\tilde{\omega}+K_{jj}^{\frac{\pi 
}{2}\left( 1\right) }}{D^{\prime }}\frac{1}{\tilde{\omega}}  \nonumber \\
K_{jj^{\prime }}^{\frac{\pi }{2}} &=&-\frac{1-\tilde{\delta}^{2}}{\cos
^{2}\psi }\frac{\tilde{\omega}^{4}+K_{jj^{\prime }}^{\frac{\pi }{2}\left(
4\right) }\tilde{\omega}^{3}+K_{jj^{\prime }}^{\frac{\pi }{2}\left( 3\right)
}\tilde{\omega}^{2}+K_{jj^{\prime }}^{\frac{\pi }{2}\left( 2\right) }\tilde{%
\omega}+K_{jj^{\prime }}^{\frac{\pi }{2}\left( 1\right) }}{D^{\prime }} 
\nonumber \\
K_{j0}^{\frac{\pi }{2}} &=&\tilde{\gamma}_{0}^{\prime }\left(
E_{eff}-1\right) \left( 1+\tilde{\delta}_{j}\right) \frac{\tilde{\omega}%
^{3}+K_{j0}^{\frac{\pi }{2}\left( 4\right) }\tilde{\omega}^{2}+K_{j0}^{\frac{%
\pi }{2}\left( 3\right) }\tilde{\omega}+K_{j0}^{\frac{\pi }{2}\left(
2\right) }}{D^{\prime }}\frac{1}{\tilde{\omega}}  \label{Kappaspi}
\end{eqnarray}
with $D^{\prime }\left( \tilde{\omega}\right) $ defined in (\ref{D-prime})$.$

At resonance the $K_{kl}^{0,\frac{\pi }{2}}$ reduce to 
\begin{eqnarray*}
K_{jj}^{0} &=&-i\left( 1+\tilde{\delta}_{j}\right) \frac{\tilde{\omega}%
^{2}-i\left( 1+\tilde{\gamma}_{0}^{\prime }-\tilde{\delta}_{j}\right) \tilde{%
\omega}-\tilde{\gamma}_{0}^{\prime }E\left( 1-\tilde{\delta}_{j}\right) }{%
D_{+}} \\
K_{jj^{\prime }}^{0} &=&i\left( 1-\tilde{\delta}^{2}\right) \frac{\tilde{%
\omega}+i\tilde{\gamma}_{0}^{\prime }\left( E-2\right) }{D_{+}} \\
K_{j0}^{0} &=&-\tilde{\gamma}_{0}^{\prime }\left( E-1\right) \left( 1+\tilde{%
\delta}_{j}\right) \frac{\tilde{\omega}-i2\left( 1-\tilde{\delta}_{j}\right) 
}{D_{+}}
\end{eqnarray*}
and 
\begin{eqnarray*}
K_{jj}^{\frac{\pi }{2}} &=&-i\left( 1+\tilde{\delta}_{j}\right) \frac{\tilde{%
\omega}^{2}-i\left( 1+\tilde{\gamma}_{0}^{\prime }-\tilde{\delta}_{j}\right) 
\tilde{\omega}-\tilde{\gamma}_{0}^{\prime }E\left( 1-\tilde{\delta}%
_{j}\right) }{D_{-}}\frac{1}{\tilde{\omega}} \\
K_{jj^{\prime }}^{\frac{\pi }{2}} &=&\left( 1-\tilde{\delta}^{2}\right) 
\frac{\tilde{\omega}-i\tilde{\gamma}_{0}^{\prime }E}{D_{-}}\frac{1}{\tilde{%
\omega}} \\
K_{j0}^{\frac{\pi }{2}} &=&-\tilde{\gamma}_{0}^{\prime }\left( E-1\right)
\left( 1+\tilde{\delta}_{j}\right) \frac{1}{D_{-}}
\end{eqnarray*}

In the adiabatic case ($\tilde{\gamma}_{0}^{\prime }\gg 1$) 
\begin{eqnarray*}
K_{jj}^{0} &=&-i\left( 1+\tilde{\delta}_{j}\right) \frac{\tilde{\omega}%
-iE\left( 1-\tilde{\delta}_{j}\right) }{\tilde{\omega}^{2}-2iE\tilde{\omega}%
-4\left( E-1\right) \left( 1-\tilde{\delta}^{2}\right) } \\
K_{jj^{\prime }}^{0} &=&-i\left( E-2\right) \left( 1-\tilde{\delta}%
^{2}\right) \frac{1}{\tilde{\omega}^{2}-2iE\tilde{\omega}-4\left( E-1\right)
\left( 1-\tilde{\delta}^{2}\right) } \\
K_{j0}^{0} &=&-i\left( E-1\right) \left( 1+\tilde{\delta}_{j}\right) \frac{%
\tilde{\omega}-i2\left( 1-\tilde{\delta}_{j}\right) }{\tilde{\omega}^{2}-2iE%
\tilde{\omega}-4\left( E-1\right) \left( 1-\tilde{\delta}^{2}\right) }
\end{eqnarray*}
and 
\begin{eqnarray*}
K_{jj}^{\frac{\pi }{2}} &=&-i\left( 1+\tilde{\delta}_{j}\right) \frac{\tilde{%
\omega}-iE\left( 1-\tilde{\delta}_{j}\right) }{\tilde{\omega}-i2E}\frac{1}{%
\tilde{\omega}} \\
K_{jj^{\prime }}^{\frac{\pi }{2}} &=&E\left( 1-\tilde{\delta}^{2}\right) 
\frac{1}{\tilde{\omega}-i2E}\frac{1}{\tilde{\omega}} \\
K_{j0}^{\frac{\pi }{2}} &=&-i\left( E-1\right) \left( 1+\tilde{\delta}%
_{j}\right) \frac{1}{\tilde{\omega}-i2E}
\end{eqnarray*}

\section{Appendix B}

The coefficients $D^{\left( j\right) }$ of the characteristic polynomial are
given by

\begin{eqnarray}
D^{\left( 5\right) } &=&-i2\left( \tilde{\gamma}_{0}^{\prime }+2\right) 
\nonumber \\
D^{\left( 4\right) } &=&-\left( \tilde{\gamma}_{0}^{\prime }+2\right) ^{2}-4%
\tilde{\gamma}_{0}^{\prime }-4\tilde{\delta}^{2}\tan ^{2}\psi -4\left(
E_{eff}-1\right) \tilde{\gamma}_{0}^{\prime }  \nonumber \\
D^{\left( 3\right) } &=&i4\tilde{\gamma}_{0}^{\prime }\left[ E_{eff}\tilde{%
\gamma}_{0}^{\prime }+2+3\left( E_{eff}-1\right) +\left( 1+E_{eff}\right) 
\tilde{\delta}^{2}\right]  \nonumber \\
D^{\left( 2\right) } &=&4\tilde{\gamma}_{0}^{\prime }\left[ \left\{ \left(
E_{eff}-1\right) ^{2}+3\left( E_{eff}-1\right) +1\right\} \tilde{\gamma}%
_{0}^{\prime }+2\left( E_{eff}-1\right) -\left\{ \left( 2+\tilde{\gamma}%
_{0}^{\prime }\right) \left( E_{eff}-1\right) -\tilde{\gamma}_{0}^{\prime
}\tan ^{2}\psi \right\} \tilde{\delta}^{2}\right]  \nonumber \\
D^{\left( 1\right) } &=&-i8\tilde{\gamma}_{0}^{\prime }E_{eff}\left(
E_{eff}-1\right) \left( 1-\tilde{\delta}^{2}\right)  \label{coefficients}
\end{eqnarray}
with $\tilde{\gamma}_{0}$ and $\tilde{\delta}$ respectively the normalized
pump mode damping and mismatch factor

\begin{center}
{\Large Acknowledgments}
\end{center}

Two of the authors (P.A. and S.S.) wish to express their thanks to the Erwin
Schr\"{o}dinger Institute in Wien for the kind hospitality offered during
the completion of this work. A.P.,C.A.,C.de L., and S.S. have participated
in this work in the frame of the project PAIS ''TWIN'' of the INFM Sect. A.

\begin{center}
{\large Figure captions}
\end{center}

\begin{enumerate}
\item  Zeros $\Omega _{1},\Omega _{2},\Omega _{4}$ of $D^{\prime }\left(
\omega \right) $ (see Eq. (\ref{D-prime})), versus pump parameter $E$ for $%
\psi =0.3$ and cavity parameters $\tilde{\gamma}_{0}^{\prime }=2,\;\tilde{%
\delta}=0.1$. $w_{1}=%
\mathop{\rm Im}%
\left[ \tilde{\Omega}_{1}\right] $ (dashed), $\tilde{\omega}_{2}=%
\mathop{\rm Re}%
\left[ \tilde{\Omega}_{2}\right] $ and $w_{2}=%
\mathop{\rm Im}%
\left[ \tilde{\Omega}_{2}\right] $ (dotted), and $\tilde{\omega}_{4}=%
\mathop{\rm Re}%
\left[ \tilde{\Omega}_{4}\right] $ and $w_{4}=%
\mathop{\rm Im}%
\left[ \tilde{\Omega}_{4}\right] $ (continuous). Notice that $%
\mathop{\rm Re}%
\left[ \tilde{\Omega}_{2,4}\right] $ vanishes for $E$ less a threshold
value, in agreement with the discussion illustrated in Fig. 2.

\item  Typical shape for $\tilde{\delta}=0.05$ of the curve separating the
points of the plane $\tilde{\gamma}_{0}^{\prime }-\,E,$ for which the three
resonance frequencies of the amplitude quadratures $\mu $ are purely
imaginary (below), from those with two complex conjugate roots (above).
Relaxation oscillations occur only for points lying above the curve.

\item  Frequency $\tilde{\omega}_{2}=%
\mathop{\rm Re}%
\left[ \tilde{\Omega}_{2}\right] \left( 
\mathop{\rm Re}%
\left[ \tilde{\Omega}_{3}\right] \right) $ (continuous) and damping
coefficient $w_{2}=%
\mathop{\rm Im}%
\left[ \tilde{\Omega}_{2,3}\right] $ (dotted), of the relaxation
oscillations at resonance ($\psi =0$), versus the pump parameter $E$ (see
Eq. (\ref{relaxation})) for cavity decay rates $\tilde{\gamma}_{0}^{\prime
}=0.5,\,2.5,\,4.5,6.5$ and $\tilde{\delta}=0$.

\item  In-phase absolute squared transfer functions $\left| K_{11}^{0}\left( 
\tilde{\omega}\right) \right| ^{2},\left| K_{12}^{0}\left( \tilde{\omega}%
\right) \right| ^{2},$ and $\left| K_{10}^{0}\left( \tilde{\omega}\right)
\right| ^{2}$ at resonance ($\psi =0$) vs normalized frequency for $\tilde{%
\gamma}_{0}^{\prime }=1$, $\tilde{\delta}=0.05$, and pump parameters $%
E=2,3,4.$

\item  Average phase commutator $\left\langle \left[ \phi _{j}\left(
0\right) ,\phi _{j}\left( \tau \right) \right] \right\rangle $ versus $\tau $
for $\tilde{\gamma}_{0}^{\prime }=4,$ ${\cal E}=1.5,4$ and $\psi =0.1.$

\item  Single mode spectra for pump cavity damping $\tilde{\gamma}%
_{0}^{\prime }=\;4$ (top),$\;0.5$ (bottom), $\tilde{\delta}=0.05$, pump
parameter $E=3$, detuning phases $\psi =0$ (solid), $0.25$ (dashed), $0.5$
(dot--dashed), and crystal losses $\varkappa _{1}/\gamma =$ $0.3$, $%
\varkappa _{0}=$ $\varkappa _{1}/3$. The spectra have been normalized to the
shot--noise--level.

\item  Effects of the detuning on the transfer of pump excess noise to beam
fluctuations. Plots parameters are as follow: $\tilde{\gamma}_{0}^{\prime
}=\;4\;$(top),$\;0.5$ (bottom), $\tilde{\delta}=0.05$, pump parameter $E=3$,
detuning phases $\psi =0$ (solid), $0.25$ (dashed), $0.5$ (dot--dashed), and
crystal losses $\varkappa _{1}/\gamma =$ $0.3$,$\varkappa _{0}=$ $\varkappa
_{1}/3$.

\item  Difference spectral density $S_{d}$ for cavity damping $\tilde{\gamma}%
_{0}^{\prime }=\;4\;$(top),$\;0.5$ (bottom), $\tilde{\delta}=0.05$, pump
parameter $E=3$, zero detunings ($\psi =0$), and crystal losses different
for the pump and signal/idler $\varkappa _{1,2}/\gamma =0.3,\,\varkappa
_{0}=\varkappa _{1}/3.$ The spectra have been normalized to the
shot--noise--level.
\end{enumerate}

\end{document}